\documentclass[twocolumn]{aastex631}
\usepackage{mathrsfs}
\usepackage{tikz}
\usetikzlibrary{positioning}
\usepackage{subfigure}
\usepackage{graphicx}
\usepackage{epstopdf}
\usepackage{overpic}
\usepackage{ulem}
\usepackage{soul}
\soulregister\citep7
\soulregister\ref7
\soulregister\citealt7
\soulregister\cite7
\usepackage{float}
\usepackage{lineno}

\makeatletter

\newcommand{\Rmnum}[1]{\expandafter\@slowromancap\romannumeral #1@}
\makeatother

\shorttitle{Simulation of AR 11429 eruption}
\shortauthors{}

\graphicspath{{./}{figures/}}

\begin{document}

\title{MHD simulation of Solar Eruption from Active Region 11429 Driven by Photospheric Velocity Field}

\author{Xinyi Wang}
\affiliation{SIGMA Weather Group, State Key Laboratory for Space Weather, National Space Science Center, Chinese Academy of Sciences,Beijing 100190, PR China}
\affiliation{College of Earth and Planetary Sciences, University of Chinese Academy of Sciences, Beijing 100049, PR China}

\author[0000-0002-7018-6862]{Chaowei Jiang}
\affiliation{Institute of Space Science and Applied Technology, Harbin Institute of Technology, Shenzhen 518055, PR China, \url{chaowei@hit.edu.cn}}

\author[0000-0001-8605-2159]{Xueshang Feng}
\affiliation{SIGMA Weather Group, State Key Laboratory for Space Weather, National Space Science Center, Chinese Academy of Sciences,Beijing 100190, PR China}
\affiliation{Institute of Space Science and Applied Technology, Harbin Institute of Technology, Shenzhen 518055, PR China, \url{chaowei@hit.edu.cn}}
%
%

\begin{abstract}
Data-driven simulation is becoming an important approach for realistically characterizing the configuration and evolution of solar active regions, revealing the onset mechanism of solar eruption events and hopefully achieving the goal of accurate space weather forecast, which is beyond the scope of any existing theoretical modelling. Here we performed a full 3D MHD simulation using the data-driven approach and followed the whole evolution process from quasi-static phase to eruption successfully for solar active region NOAA 11429. The MHD system was driven at the bottom boundary by photospheric velocity field, which is derived by the DAVE4VM method from the observed vector magnetograms. The simulation shows that a magnetic flux rope was generated by persistent photospheric flow before the flare onset and then triggered to erupt by torus instability. Our simulation demonstrates a high degree of consistency with observations in the pre-eruption magnetic structure, the time scale of quasi-static stage, the pattern of flare ribbons as well as the time evolution of magnetic energy injection and total unsigned magnetic flux. We further found that an eruption can also be initiated in the simulation as driven by only the horizontal components of photospheric flow, but a comparison of the different simulations indicates that the vertical flow at the bottom boundary is necessary in reproducing more realistically these observed features, emphasizing the importance of flux emergence during the development of this AR.
\end{abstract}


\section{Introduction}
As driven by solar eruptions, the solar-terrestrial environment often experiences variations, which are known as space weather, and forecasting the space weather precisely is not only an important scientific topic but can also avoid damage of the sensitive on-ground and space-based critical infrastructures. Though many theoretical models have been proposed and significant process has been made in understanding the triggering mechanism of solar eruptions \citep{2000JGR10523153F, 2011LRSP....8....1C,2013AdSpR..51.1967S,2014masu.book.....P}, reproducing the whole life-span from quasi-static stage to eruption using numerical models constrained and driven by observed vector magnetogram possess unprecedented capabilities in revealing the onset mechanism of the real eruption events, and can potentially be used for accurate space weather forecast \citep{JIANG2022100236,2022-022}.

Previous study reproduced the whole process of energy accumulation and release successfully \citep{2016NatCo...711522J}, showing an MHD system can be driven to erupt by inputting time series of vector magnetograms at the bottom boundary (B-driven). There are also other data-driven models, in which the evolution of MHD system is driven by the electric field (E-driven, e.g., \citealt{2012ApJ...757..147C,Hayashi_2018,2019SoPh..294...41P,2019A&A...628A.114P}) or the velocity field (V-driven, e.g., \citealt{2019ApJ...871L..28H,2019ApJ...870L..21G,2019A&A...626A..91L,2020ApJ...892....9H,2021NatCo..12.2734Z}) on the photosphere (bottom boundary). Though the B-driven method can fully match the magnetogram, it will introduce considerable errors of magnetic divergence from the bottom boundary. This shortage will vanish in E-driven model, however, deriving both the induction and potential components of the electric field on the photosphere is not an easy task \citep{2010ApJ...715..242F,2012SoPh..277..153F,2015SpWea..13..369F,2020ApJS..248....2F}, and the photospheric flow also needs to be properly set to follow the Ohm's law. In most of the theoretical models of solar eruption, the key structure in favor of eruption is assumed to be formed through the movement of the footpoints of magnetic field lines \citep{1980IAUS...91..207M,1992LNP...399...69M,1999ApJ...510..485A,2000JGR...105.2375L,2021NatAs...5.1126J}, which is driven by the horizontal flow, and the vertical component of the photospheric flow will be responsible for the flux emergence process. Therefore, with the photospheric velocity field determined, the photospheric magnetic field can be generated self-consistently. Furthermore, in the V-driven approach, there is no need to solve the complex momentum equation at the bottom boundary (which is the most time-consuming part in solving MHD equations). Due to these advantages, the V-driven method has attracted many previous studies to focus on this topic. For example, with the velocity field derived by DAVE4VM method \citep{2008ApJ...683.1134S}, \cite{2019ApJ...871L..28H} used the projected normal characteristics method to update the physical variables other than velocity at the bottom boundary. However, the total magnetic energy kept almost the same level of that of the initial state without obvious magnetic energy injection. \cite{2021FrP.....9..224J} updated the magnetic field by solving directly the magnetic induction equation at the bottom boundary and their model can inject the magnetic energy from bottom boundary successfully. \cite{2020ApJ...892....9H} drove the magnetic evolution by inputting the DAVE4VM velocity field and vector magnetogram simultaneously. The formation process of an magnetic flux rope (MFR) was obtained in their model. Unfortunately, these simulations didn't drive the system to erupt. The only work we know that obtained an eruption by the velocity field derived from observation was shown in \cite{2021ApJ...909..155K}. The velocity field was derived from the electric field on the photosphere using Ohm's law and then was input at the bottom boundary in their zero-beta model to drive the system to erupt. Two eruptions they produced were identified from the evolution curves of the magnetic and kinetic energies, however, the kinetic energy showed an overall increase without obvious quasi-static evolution during the pre-eruption stage. Before the first eruption, the kinetic energy was comparable with its peak during the first eruption and the amount of the release of the magnetic energy was also too small, which didn't show the feature of a typical eruption event, i.e., a large amount of magnetic energy was converted into kinetic energy impulsively. As described above, the whole energy accumulation process from quasi-static evolution (of typically tens of hours) to impulsive eruption has not been realized in a self-consistent way using the V-driven MHD model, and this is one of the motivations of this work.

In this Letter, we applied a V-driven model to investigate the evolution and eruption of a well-studied active region (AR) NOAA 11429. Previous studies found persistent shearing flow and flux cancellation near the main polarity inversion line (PIL) of this AR \citep{8210419,2017RAA....17...81Z}, which were suggested to be responsible for the eruptions on 2012 March 7, 9 and 10 \citep{2018ApJ...860...35D,2020ApJ...901...40D}. An analysis of the MFR reconstructed from vector magnetograms using the nonlinear force-free field (NLFFF) model suggests that the homologous eruptions are triggered by torus instability (TI) of the MFR \citep{2015ApJ...809...34C}. \cite{2021ApJ...910...40Z} suggested the helical kink instability may also take effect. Nevertheless, whether these mechanisms were at work requires to be further studied using dynamic modeling of this AR evolution and eruption, which is absent in all the previous studies and is the other motivation of this work. In our simulation, the dynamic process from the beginning of 2012 March 4 to the eruption on March 5 in this AR was reproduced self-consistently as driven by the photospheric velocity derived from vector magnetogram using DAVE4VM method. Our simulation shows that an MFR was generated near the main PIL before the flare onset and the initiation of this eruption event depended mainly on TI of the preformed MFR.


\section{Data and Model}\label{Data and Model}
\begin{figure}[htbp]
	\centering  
	\includegraphics[scale=0.8]{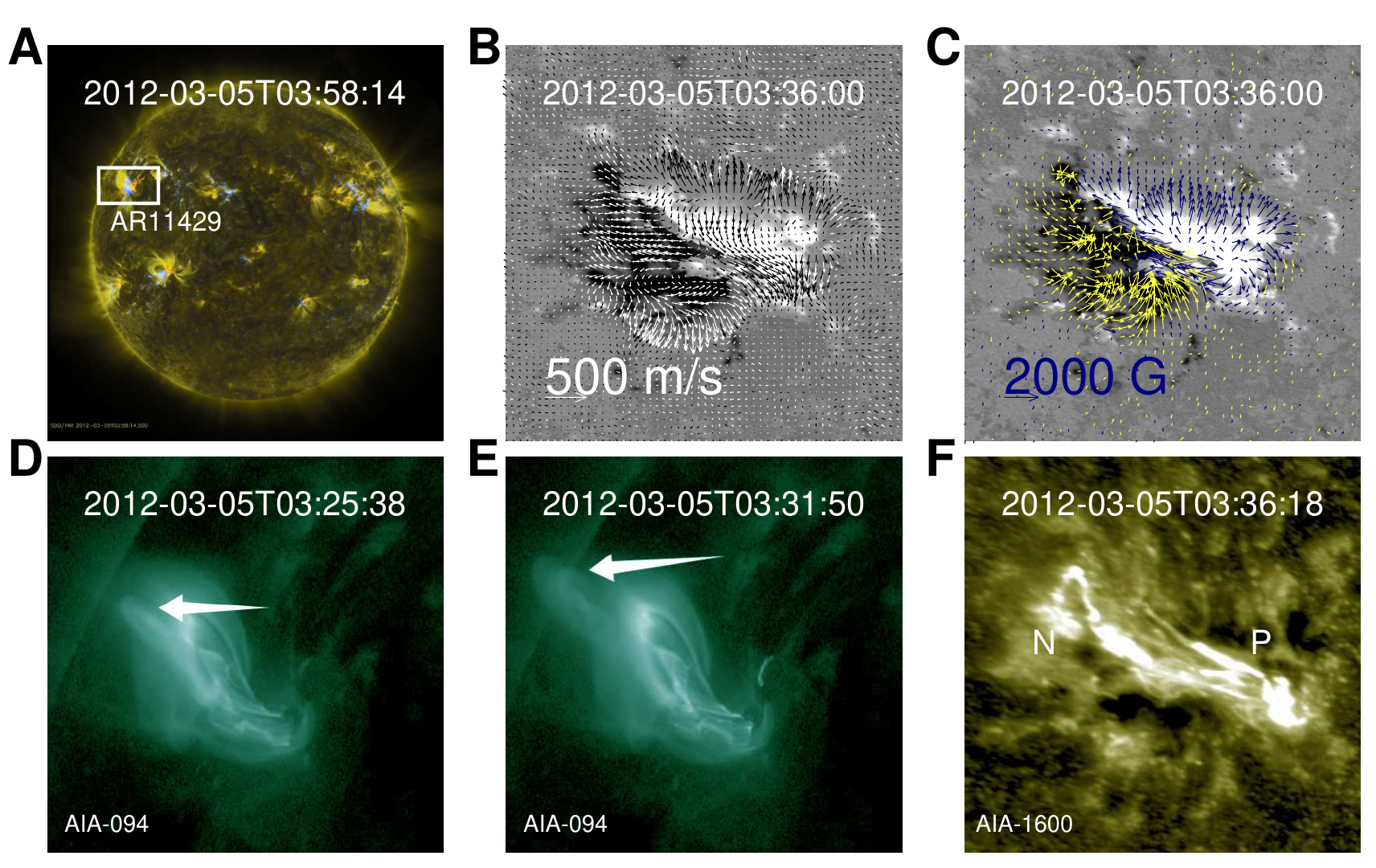}
	\caption{($\mathbf{A}$): A full-disk composite image of SDO AIA 171 $\rm \AA$ and HMI magnetogram. The white box shows the AR 11429. ($\mathbf{B}$): The vectors denote the time average horizontal velocity from 2012 March 4 00:00 UT to March 5 03:36 UT on the photosphere. The background shows the magnetic flux distribution, in which black and white represents the negative and positive polarity respectively. ($\mathbf{C}$): The vectors denote the transverse magnetic field on the photosphere. ($\mathbf{D}$): Pre-flare image of AIA 94$\rm \AA$. The white arrow shows the erupting structure. ($\mathbf{E}$): Same as $\mathbf{D}$, but at a time closer to eruption before the flare onset. ($\mathbf{F}$): The first image of flare ribbons observed in AIA 1600$\rm \AA$. Letter P labels the positive ribbon and N labels the negative ribbon.}
	\label{fig1}
\end{figure}

AR 11429 showed a complex $\beta\gamma\delta$ configuration and is very flare-productive, which has produced 3 X-class flares from 2012 March 5 to 7. It first appeared on the eastern solar limb on March 4 and was located on the eastern part of the solar disk before March 8. During this period, the AR kept developing as characterized by the increasing total unsigned magnetic flux, indicating the obvious flux emergence by vertical flow on the photosphere. Since it was the first X-class flare of this AR, here we focus on the initiation process of the X1.1 flare (which is accompanied with a halo CME moving at a speed of 1531 $\rm km~s^{-1}$) around 04:00 UT on March 5 as shown in the white box in Figure~\ref{fig1}A. Before the flare onset, there was persistent shearing flow near the main PIL (Figure~\ref{fig1}B) and as a result, the horizontal magnetic field there was highly sheared (Figure~\ref{fig1}C), which indicates that a large amount of free magnetic energy is stored ready for an eruption. An hot loop first erupted away as shown in AIA 94 $\rm \AA$ \citep{2012SoPh..275...17L} at around 03:31 UT (Figure~\ref{fig1}D and E) and after that a pair of hook shape flare ribbons appeared near the main PIL in AIA 1600 $\rm \AA$ at 03:36 UT (Figure~\ref{fig1}F), i.e., the flare event started.

 To understand the formation of the pre-eruptive coronal magnetic field and the triggering mechanism of this eruption, we used the DARE-MHD model \citep{2016NatCo...711522J} to study the dynamic evolution of this AR. For saving the computational time, the strength of magnetic field from the magnetograms were reduced by a factor of 25 before being input into our code. The initial plasma density was set as a hydrostatic isothermal model with a fixed temperature as $T=1\times 10^{6}$ K and a modified solar gravity \citep{2021NatAs...5.1126J} to get a plasma background that mimics the real environment in the solar corona basing on two key parameters, the plasma $\beta$ and the $\rm Alfv \acute{e} n$ speed (in particular, the minimum $\beta$ is $6.3\times 10^{-4}$ and maximum $\rm Alfv \acute{e} n$ speed $V_{\rm A}\sim 4800$~km~s$^{-1}$ in the final equilibrium we obtained below). We chose to use the magnetograms of HMI SHARP data set \citep{2012SoPh..275..229S,2014SoPh..289.3549B} at 00:00UT on March 4 to reconstruct the initial magnetic field since there were no obvious MFR at that time. This magnetogram was smoothed first using Gaussian smoothing with FWHM of 6 pixels and a NLFFF model from this smoothed magnetogram was extrapolated by our CESE-NLFFF-MHD code \citep{2012ApJ...749..135J,2013ApJ...769..144J}. Since the code (like many other NLFFF codes) does not give a perfect force-free solution but with residual Lorentz forces, we input the extrapolated field into the DARE-MHD model, along with the initial background plasma,  to let the MHD system relax until the kinetic and magnetic energies were almost unchanged, i.e., an MHD equilibrium was obtained, and the initial state was ready.

 At the bottom boundary, we solve the magnetic induction equation to update the magnetic field with the velocity field derived by DAVE4VM method to update the magnetic field on the photosphere. The DAVE4VM velocity was strengthened by a factor of 13.7 (determined by the ratio of the time series magnetograms' original time cadence as 720 seconds to the time cadence in our simulation as $0.5 \times 105$ seconds, i.e., $\frac{720}{0.5\times 105}$) to speed up our simulation and thus the time scale of quasi-static evolution prior to the eruption onset is shorten by the same times. At the side and top boundaries, all the variables are extrapolated from the neighboring inner points with zero gradient along the normal direction of the boundary surface and the normal component of magnetic field is further modified by the divergence-free condition. The Powell-source terms and diffusion control terms was used to deal with the divergence error of the magnetic field as described in \cite{2010SoPh..267..463J}. We set the computational domain sufficiently large as $[-368,368]$ Mm in both $x$ and $y$ direction and $[0,736]$ Mm in $z$ direction with grid resolution varies from $1^{\prime \prime}$ to $8^{\prime \prime}$ using adaptive mesh refinement. The highest resolution is used mainly for the regions with strong magnetic gradients and current density, in particular, the current sheets (CSs). Explicit value of magnetic resistivity was not used in our simulation and the magnetic reconnection was controlled by the resistivity of the numerical method only, which mimicked the low-resistivity plasma better. As the total unsigned flux of this AR kept increasing before the flare onset, we energized the MHD system by full 3D DAVE4VM velocity field ($v^{D}_{x}, v^{D}_{y}, v^{D}_{z}$) (will be referred to as V3D simulation) and horizontal component ($v^{D}_{x}, v^{D}_{y}, 0$) (V2D) respectively, and a comparison of the results for these two simulations will show the importance of flux emergence through the vertical flow on the photosphere.
\section{Results}\label{Results}
\subsection{Overall Process}\label{Overall Process}
The evolution curves of the total magnetic and kinetic energies in the computational domain as well as the total unsigned magnetic flux at the bottom boundary are shown in Figure~\ref{flux_energy}. For the `V3D' simulation, as driven by the time-series velocity field (V3D) for a time duration of 150 minutes, the total magnetic energy in our simulation model experienced an overall increase firstly and then a rapid decrease, which is associated with an eruption event. The eruption can be identified from the energy evolution with onset time $t_{\rm E,V3D}=120$ minutes.

At the very beginning from $t=0$ to $t=22$ minutes, the magnetic energy injection curve (black dashed line, which is computed by the time integration of the total Poynting flux of the `V3D' simulation at the bottom surface) matches well with the solid `V3D' curve (the magnetic energy increase of the `V3D' simulation) and `OB' curve (the magnetic energy injection computed by DAVE4VM velocity and magnetograms) in Figure~\ref{flux_energy}A. However, in the time duration of $t\in [97, 122]$ minutes, the total magnetic energy (blue solid line) is higher than the magnetic energy injection. Such an unphysical mismatch of the inputted energy from the boundary and the cumulative energy in the volume is likely owing to the insufficient resolution for the bottom boundary, since the magnetic field at the bottom boundary has accumulated a very large gradient in this phase (see Discussions). The kinetic energy keeps a very low value of around $10^{-3} E_{\rm p_{0}}$ (which is the potential field energy corresponding to the magnetic field at t=0) before the major eruption begins at $t_{\rm E,V3D}=120$ minutes when the magnetic energy reaches about 1.85 $E_{\rm p_{0}}$. The magnetic energy decreases to about 1.7 $E_{\rm p_{0}}$ at the peak of the total kinetic energy (i.e., $E_{\rm k} = 5.6 \times 10^{-2} E_{\rm p_{0}}$), and keeps decreasing to 1.45 $E_{\rm p_{0}}$ in total through this eruption (0.4 $E_{\rm p_{0}}$ free energy loss). That is, about one third of the magnetic energy loss has been converted to kinetic energy in 10 minutes. If multiplied by a factor of 13.7 determined by the rate of speeding up in our velocity-driven simulation, the quasi-static evolution time of `V3D' run is 27.4 hours. This time scale is very close to the observation one which is 27.6 hours.

We have also driven the simulation by the horizontal velocity (V2D) and found that it can also produce an eruption with rather similar onset time. However, comparing the different curves of magnetic energy evolution in Figure~\ref{flux_energy}A and the curves of total unsigned magnetic flux in Figure~\ref{flux_energy}B respectively, the blue solid lines labeled by `V2D' are obvious lower than those from the 'V3D' simulation. This clearly shows that though horizontal velocity can also drive the field to erupt (with a delayed onset time for about 6 minutes compared with `V3D'), the vertical velocity $v^{D}_{z}$ is necessary in accounting for the larger increase of the total magnetic energy and total unsigned flux as shown in observations, therefore leading to a stronger eruption. The evolution of the simulated magnetic energy, total unsigned flux and the time scale of the `V3D' run before the major eruption are more consistent with observations, showing the importance of the vertical photospheric plasma flow in the numerical modeling of solar eruptions. Our simulated magnetic structure (the first panel in Figure~\ref{fig3}E) has reasonable consistency with observations in the pre-eruption image of AIA 171 $\rm \AA$ (the second panel in Figure~\ref{fig3}E), and the synthetic image of coronal emission from current density of our simulation (the last panel in Figure~\ref{fig3}E) is reasonable consistent with the image of AIA 131 $\rm \AA$ (the third panel in Figure~\ref{fig3}E). Also the quasi separatrix layers (QSLs), where the magnetic reconnection is most likely to take place and thus represent the position of the flare ribbons \citep{2002JGRA..107.1164T,2016ApJ...818..148L}, at the bottom boundary (Figure~\ref{fig4}F) has approximately the same patterns as the flare ribbons in AIA 1600 $\rm \AA$ (Figure~\ref{fig4}E). These results confirm the validity of our V-driven DARE-MHD model as well as the DAVE4VM method.

\subsection{Eruption Initiation Mechanism}\label{trigger}
Since the actual velocity field must contain $v_{z}$, here we analyzed the `V3D' run to study the onset mechanism of this eruption. As we can see, a group of twisted field lines (represented by the blue solid lines in Figure~\ref{fig3}A and B) formed and was embedded in the surrounding shear arcades, which is similar to an MFR in morphology. In addition, the ejection of the hot loop (also called hot channel) as observed in AIA 94 $\rm \AA$ (shown by the white arrows in Figure~\ref{fig1}D and E) also suggested the existence of an MFR \citep{2017ScChD..60.1383C} before the flare onset.

To identify the formation of MFR before the eruption, we calculated the QSLs and the twist number \citep{2006JPhA...39.8321B} of our simulated magnetic fields, which are given in Figure~\ref{fig3}D, Figure~\ref{fig5}A and B, respectively. As shown in the third panel of Figure~\ref{fig3}D, a strong QSL appeared near the core field and grew up to be a QSL ring in the last panel, which separated the MFR and the background magnetic field, thus representing the existence of an MFR in topology. The QSL ring intersects itself below the MFR, forming a typical hyperbolic flux tube (HFT), where the CS developed and magnetic reconnection took place subsequently to further drive the eruption \citep{Jiang_2018}.  The isosurface of $T_{w}=-1$, which represents the position and shape of the MFR, are shown in Figure~\ref{fig5}A and B respectively. It became larger and higher, illustrating more magnetic flux were twisted as driven by persistent photospheric flow. However, the isosurface of $T_{w}=-2$, which is the lower threshold of kink instability (KI) according to the statistics of \cite{2019ApJ...884...73D}, is barely visible and KI may take little effect.

Among different triggering mechanisms, TI is considered as an efficient way of MFR eruption. In Figure~\ref{fig5}D, we plot the key controlling parameter that determines the onset of TI, i.e., the decay index of the strapping field. The variation of the decay index $n$ was calculated along the $z$ direction of the overlying field at $t=119$ minutes before the eruption onset. Since the potential field is not always the good approximation of the strapping field (especially when the strapping field is highly sheared), we calculated the decay index of our simulated field instead and plotted the decay index of the corresponding potential field at the same time for comparison. The critical height (above which $n>1.5$) of the simulated field is located at 60 Mm as labeled by the vertical black dashed line in Figure~\ref{fig5}D. When the MFR was just formed, it is small and it's apex was at a low height (about 40 Mm in Figure~\ref{fig5}A). About 14 minutes later, it had grown up to be a huge twist structure and entered the unstable zone (above 60 Mm as shown in Figure~\ref{fig5}B), after which it erupted violently.

The formation and the evolution of the MFR, i.e., there was an preformed MFR and the MFR erupted after it entered the unstable zone, strongly suggest that TI is the triggering mechanism of this eruption. To further test this assumption, we used the V3D-driven data at $t=100$ minutes as the initial condition and ran our code without bottom velocity driven (by setting all the three velocity components as zero at the bottom boundary, thus referred to as `V0D') to see whether the magnetic field will erupt. The evolution of energies of this `V0D' run are shown as the red solid lines in Figure~\ref{flux_energy}A. During the time duration of $t\in [100,124]$ minutes, the magnetic energy kept almost unchanged, while the toroidal flux of the MFR (defined as $\int_{s}B_{z}ds$, where $s$ denotes the region of $T_{w}<-1$ and $B_{z}<0$ at the bottom boundary) kept increasing as shown in Figure~\ref{fig5}C. In `V3D' simulation, the MFR (the isosurface of $T_{w}=-1$) became larger and higher than the critical height before the eruption onset (Figure~\ref{fig5}A and B) as the toroidal flux increasing (Figure~\ref{fig5}C), after which TI took effect and finally led to a strong eruption. Similarly, the toroidal flux of `V0D' simulation increased before and decreased after the eruption (Figure~\ref{fig5}C), showing the slow reconnection can took place spontaneously without velocity driving and made the MFR larger until the eruption started at $t_{\rm E,V0D}=124$ minutes as identified from the energy evolution curve. Since the current density was weaker and the current layer was thicker (as shown in the third panel in Figure~3C) than a true CS (the first panel in Figure~4C), the magnetic gradient was lower in the current layer than in the CS, and thus the diffusivity was relatively uniform, which only allowed slow reconnection (i.e., a slow dissipation of the current) to take place without impulsive energy release (i.e., not a Pescheck-type reconnection, \citealt{1994ApJ...436L.197Y}). The eruption of this `V0D' run has the similar onset time and strength with the `V3D' run, which further indicates that after $t=100$ minutes in `V3D' run, the velocity field on the bottom boundary is not necessary and the instability is sufficient to trigger the eruption. It follows the basic developing stage of TI, i.e., the rising MFR stretched the overlying field and consequently the flare CS formed below the MFR. Since we didn't use any explicit value of magnetic resistivity, and since the CS formed in a dynamic way, the width of CS could become very thin to trigger the fast reconnection easily as pointed out by \cite{2021NatAs...5.1126J}, which further drives the eruption. Based on the analysis of our simulation and the supplementary numerical experiment (`V0D' simulation), along with the consistency between our simulation results and observations that have been shown above, we conclude that TI is the initiation mechanism of the X1.1 flare in AR 11429 on 2012 March 5.

\section{Conclusions and Discussions}\label{Conclusion}
We have carried out a full 3D MHD simulation of an X-class flare eruption event on 2012 March 5 in AR 11429 using the V-driven DARE-MHD model. An MHD equilibrium was obtained by relaxing the NLFFF reconstructed by the CESE-MHD-NLFFF code and was set as the initial condition. Then the initial state was driven to evolve by the DAVE4VM velocity field on the bottom boundary of our simulation box. The analysis of the quasi-static evolution stage before the eruption onset shows the gradual formation of an MFR above the main PIL. When the MFR first appeared, it was relatively low and then it grew up into the torus unstable region, after which TI was triggered. Then the energy conversion process was accomplished by reconnection in the CS that is formed due to the stretching effect of the erupting MFR. The images of SDO observation and the important physical quantities computed from the magnetograms are reasonably consistent with our result in the pre-flare magnetic structure, the morphology of flare ribbons, evolution curves of the magnetic and kinetic energies as well as the total unsigned magnetic flux. The time duration of the quasi-static evolution process before the simulated eruption is also very close to the actual time scale before the flare onset.

\begin{figure}[htbp]
	\centering  
	\includegraphics[scale=0.8]{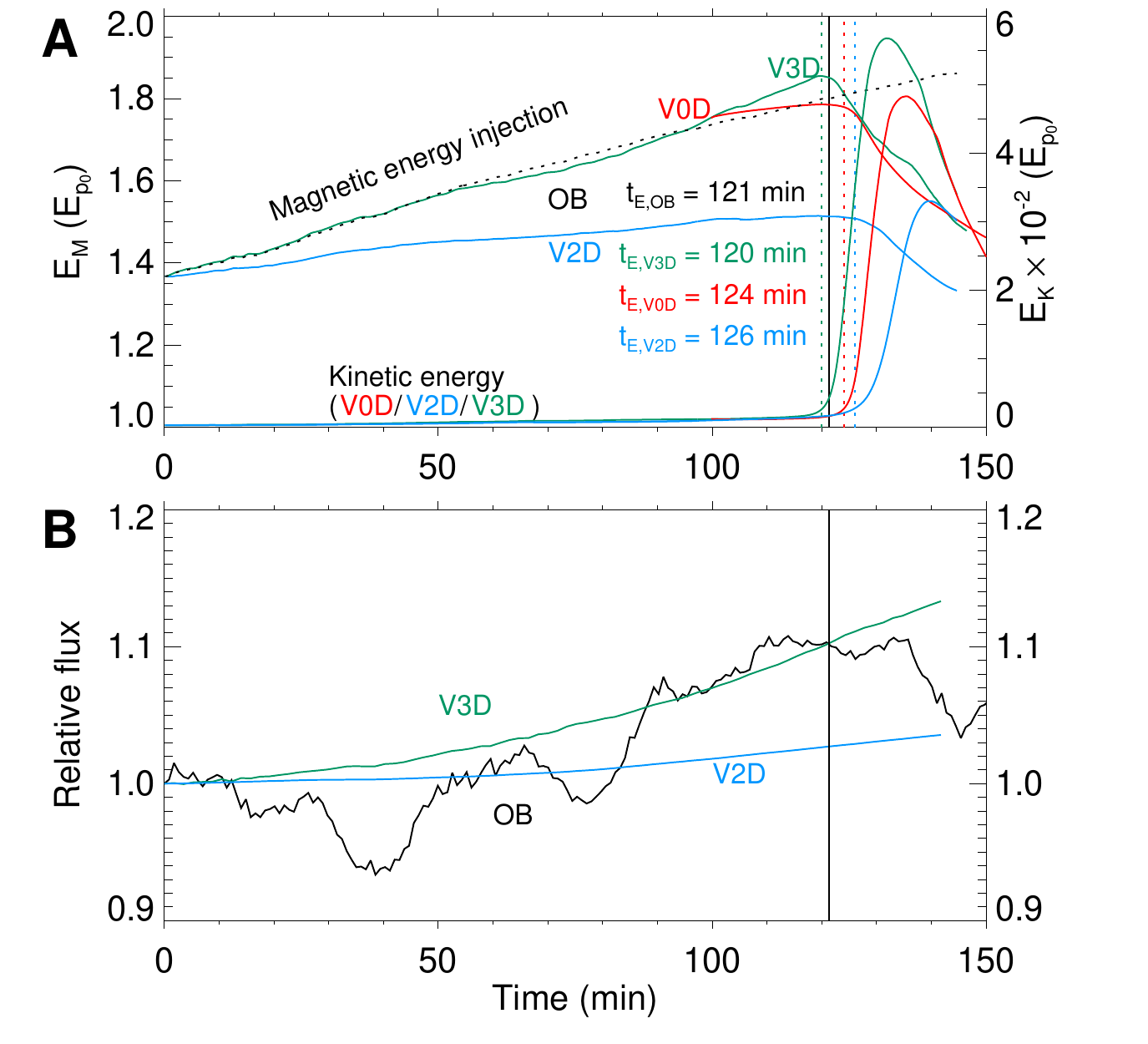}
	\caption{($\mathbf{A}$): Magnetic and kinetic energy evolution during the whole process. The unit of $x$-axis is in minutes. The solid curves in different colors denote the corresponding evolution of different energies and the vertical dashed lines in different colors denote the eruption onset time of different runs. The vertical black solid line shows the flare onset time in observations. The black curve `OB' denotes the magnetic energy injection computed by DAVE4VM velocity and magnetograms, thus representing the actual magnetic energy evolution. In our simulation, magnetic energy of the potential field of the initial magnetogram is $E_{\rm p_{0}}=2.52\times 10^{30}$ erg (which is a fixed value and is used for normalization here). This value should be multiplied by a factor of 625=$25^{2}$ if scaling to the realistic value, thus being $1.57\times 10^{33}$ erg. ($\mathbf{B}$): Time evolution of the total unsigned magnetic flux in our simulations. The `OB' curve is computed by the magnetograms and the other labels `V3D', `V2D' and the vertical black solid line have the same meanings as in $\mathbf{A}$. All curves are normalized by their initial value at $t=0$. The corresponding animation is attached for the V3D run, and it starts at $t=0$ and ends at $t=147$ minutes in simulation time, which corresponds to 33.6 hours in real-time duration. Panel $\mathbf{a}$ in the animation corresponds to rows $\mathbf{A}$ in Figure~\ref{fig3} and \ref{fig4}. Panel $\mathbf{b}$ corresponds to the rows $\mathbf{C}$ in Figure~\ref{fig3} and \ref{fig4} in time series. Panel $\mathbf{c}$ shows the corresponding evolution as in the row $\mathbf{D}$ in Figure~\ref{fig4}, and panel $\mathbf{d}$ represents the evolution of QSLs as in Figure~\ref{fig4}F. We only plotted the kinetic energy in panel $\mathbf{e}$, with the vertical solid line showing the current simulation time; the magnetic flux evolution is not included in the animation. The cadence between each snapshot used in the animation is 210 s in the quasi-static period ($t\in [0, 119]$ minutes) and 21 s in the eruption period ($t\in (119, 147]$ minutes) in simulation time, respectively.}
	\label{flux_energy}
\end{figure}

\begin{figure}[htbp]
	\centering  
	\includegraphics[scale=0.6]{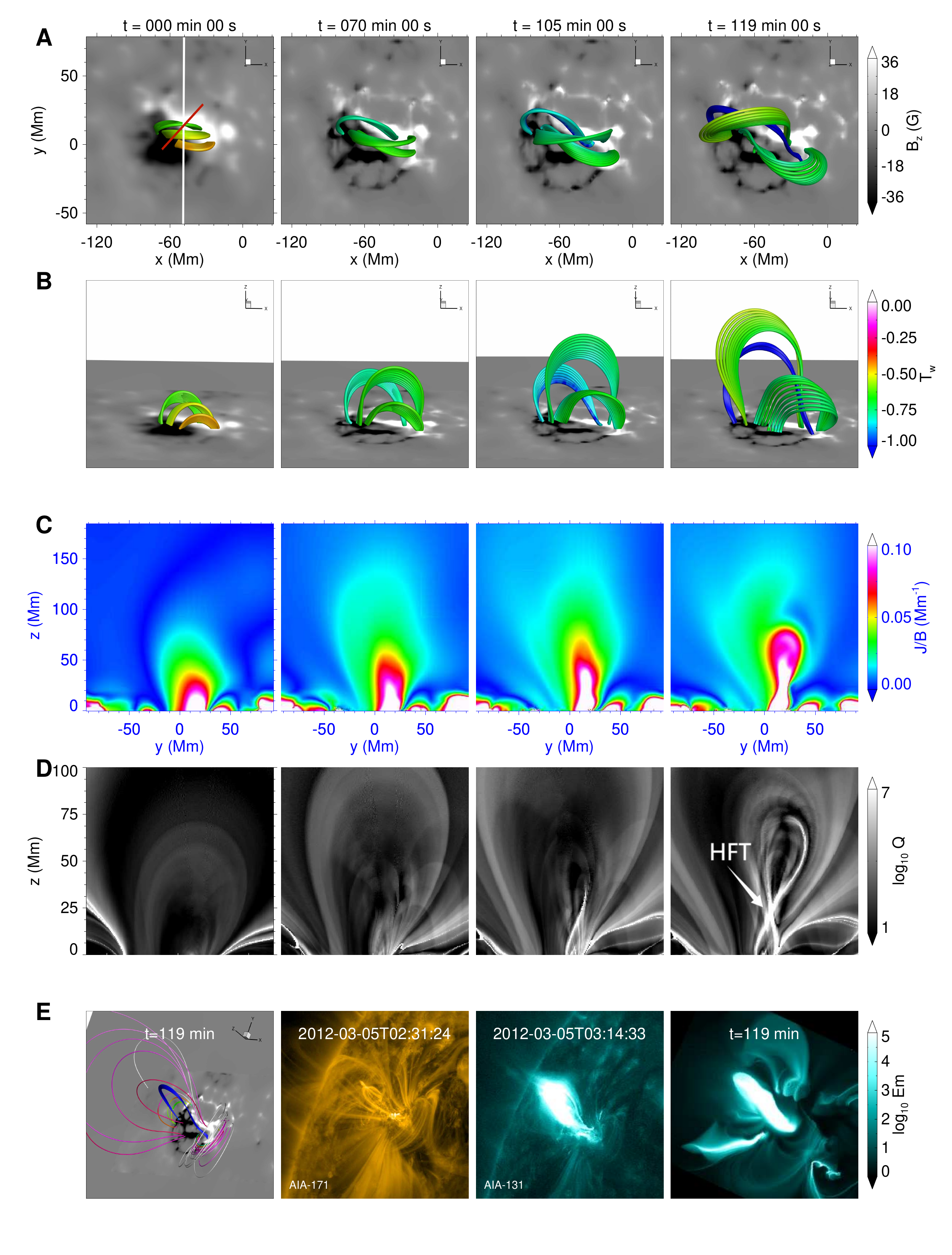}
	\caption{The magnetic evolution during the pre-eruption stage. ($\mathbf{A}$): Top view of 3 bunches of field lines with the footpoints approximately following the movement of the magnetic flux. The white solid line denotes the position of the slices in $\mathbf{C}$ and in Figure~\ref{fig4}C and D. The red solid line segment shows the position of the slices in $\mathbf{D}$, which is almost perpendicular to the PIL and at the middle of MFR. ($\mathbf{B}$): Side view of the same 3D magnetic field lines as in $\mathbf{A}$. ($\mathbf{C}$): Vertical cross section of the current layer near the main PIL. The position of these slices are labeled by the white solid line in $\mathbf{A}$. ($\mathbf{D}$): Magnetic squashing factor $Q$ on the same cross section of $\mathbf{C}$ and the region with high $Q$ denotes the QSLs. The position of these slices are labeled by the red solid line segment in $\mathbf{A}$. ($\mathbf{E}$): The first 2 panels show the comparison between our simulation and observations in AIA 171 $\rm \AA$. The last 2 panels show comparison between the synthetic image of coronal emission from current density of our simulation and observations in AIA 131 $\rm \AA$.  Row $\mathbf{A}$ is shown at panel $\mathbf{a}$, and rows $\mathbf{C}$ is shown at panel $\mathbf{c}$ in the animation of Figure~\ref{flux_energy}, respectively.}
	\label{fig3}
\end{figure}

\begin{figure}[htbp]
	\centering  
	\includegraphics[scale=0.55]{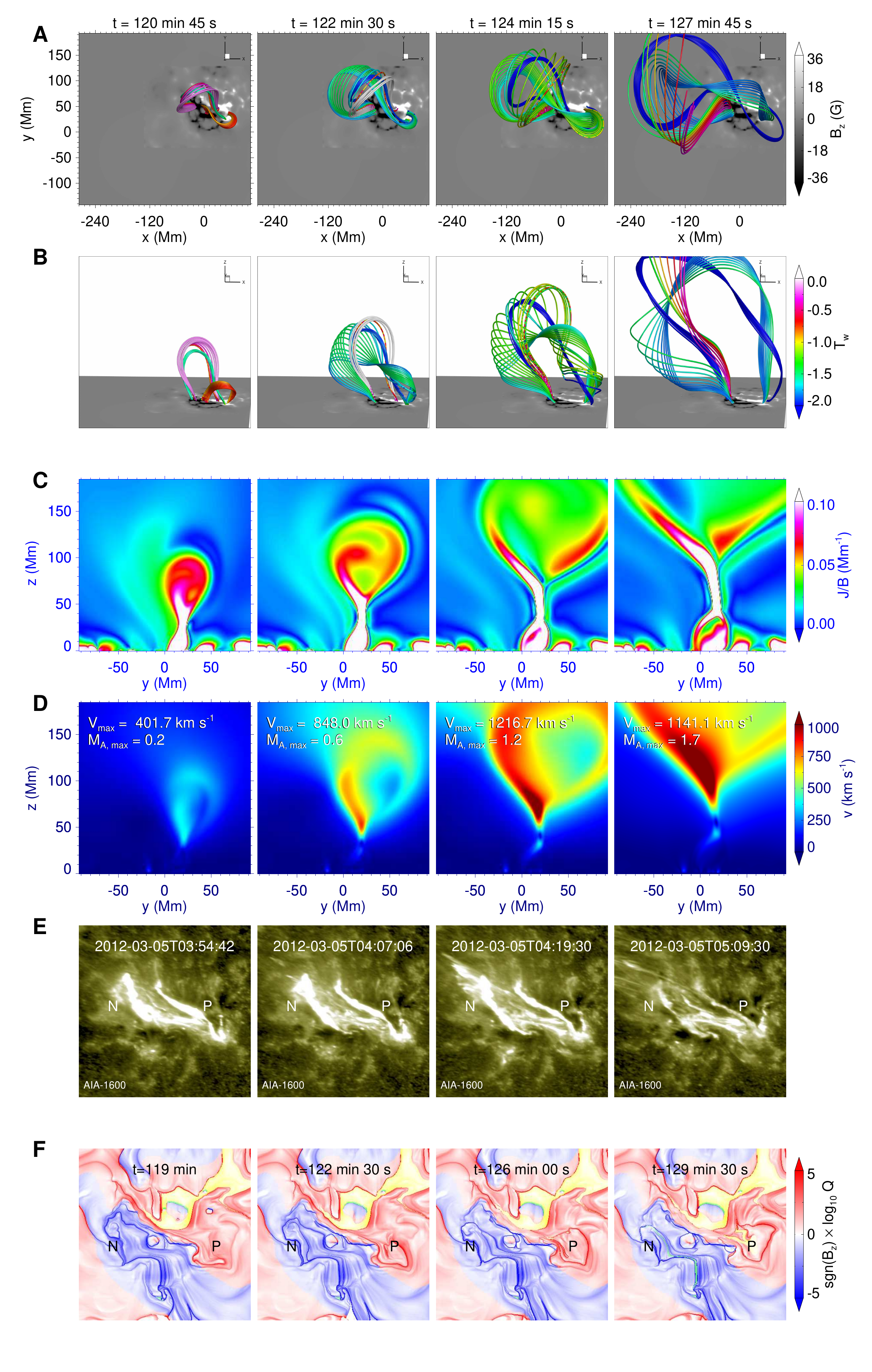}
	\caption{The magnetic evolution of the eruption stage. ($\mathbf{A}$), ($\mathbf{B}$) and ($\mathbf{C}$): All settings are the same as in Figure~\ref{fig3}A, B and C, except the footpoints of the magnetic field lines are fixed and the current layer in Figure~\ref{fig3}C turns into a current sheet here. ($\mathbf{D}$): Outflows at the position of current sheet. ($\mathbf{E}$): Projection corrected images of the flare ribbons observed in AIA 1600 $\rm \AA$. The letter P denotes the positive ribbon and N denotes the negative ribbon. ($\mathbf{F}$): Evolution of the bottom QSL of our simulation, where sgn$(B_{z})$ denotes the sign of $B_{z}$. The letter P denotes the positive QSL and N denotes the negative QSL. The slices in $\mathbf{C}$ and $\mathbf{D}$ are located at the position labeled by the white solid line in the first panel of Figure~\ref{fig3}A.  Row $\mathbf{A}$ is shown at panel $\mathbf{a}$, and rows $\mathbf{C}$ and $\mathbf{D}$ are shown at panel $\mathbf{c}$ and panel $\mathbf{d}$ in the animation of Figure~\ref{flux_energy}, respectively. The QSL evolution in row $\mathbf{F}$ is shown at panel $\mathbf{b}$.}
	\label{fig4}
\end{figure}

\begin{figure}[htbp]
	\centering  
	\includegraphics[scale=0.8]{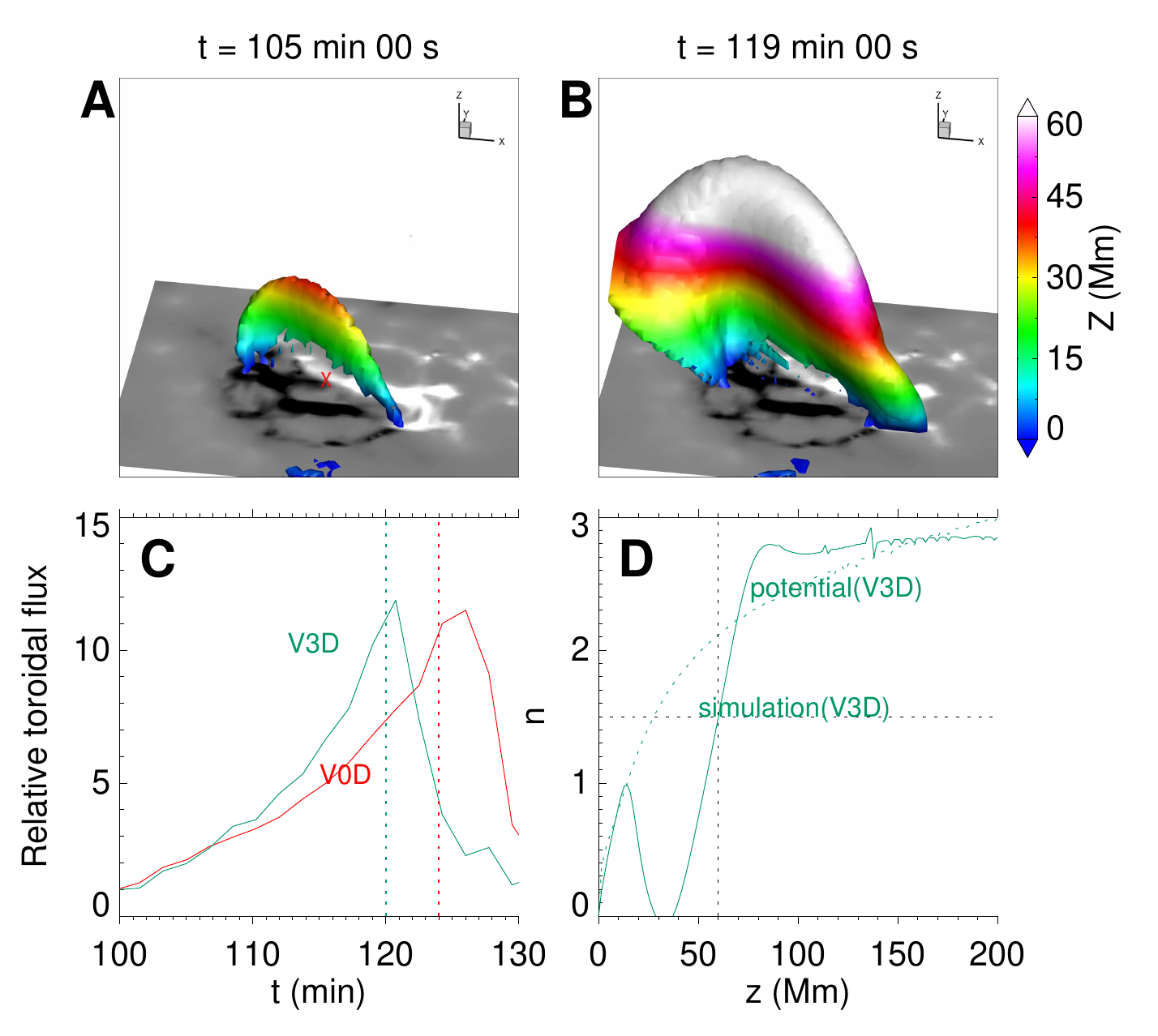}
	\caption{($\mathbf{A}$): The isosurface of $T_{w}=-1$ and $T_{w}$ is the twist number of the magnetic field in `V3D' simulation. The red `X' labels the position where we calculate the decay index. ($\mathbf{B}$): Same as $\mathbf{A}$ but at a different time. ($\mathbf{C}$): The green solid curve denotes the relative toroidal flux of the MFR in `V3D' simulation and the red solid curve in `V0D' simulation with both curves normalized by their initial values respectively. Two vertical dashed lines in different colors denotes the onset time of `V0D' and `V3D' simulations, respectively. The toroidal flux is defined as $\int_{s} B_{z}ds$, where `s' is the region of $T_{w}<-1$ and $B_{z}<0$ at the bottom boundary. ($\mathbf{D}$): The decay index $n$ of the magnetic field in `V3D' simulation (solid curve) and the corresponding potential field (dashed curve) at the same time of $t=119$ minutes. The horizontal dashed line denotes the critical value of $n=1.5$ and the vertical dashed line denotes the critical height, above which $n>1.5$.}
	\label{fig5}
\end{figure}

Nevertheless, our simulation does not reproduce accurately the evolution of magnetic field as shown in the observed magnetograms. As can be seen in the last panel of Figure~\ref{fig3}A, part of the magnetic flux was transported to and concentrated at the edge of the AR, while this pileup was not observed in HMI magnetograms. One likely reason for this flux pileup is that, the DAVE4VM method only solves the normal component of the magnetic induction equation in the least-square sense \citep{2019SoPh..294...84L}, which may not reproduce the velocity precisely at every point and can only be used to approximate the overall distribution and evolution of magnetic flux in the AR. In addition, the magnetic flux on the photosphere should be dispersed and dissipated by granular and supergranular convection \citep{1989Sci...245..712W} as well as small scale turbulent diffusion in practical situation. Therefore, without dealing properly with these effects, the flux pileup will be more obvious than observations and as a result the magnetic energy injection rate of `V3D' run will be overall higher than the `OB' one as shown in Figure~\ref{flux_energy}A (comparing the black dashed line and the black solid line). This unrealistic flux pileup may also contribute to the overshoot of the magnetic energy increase (Figure~\ref{flux_energy}A, the mismatch of the black dashed line and the green line before the eruption), because it results in a very large magnetic gradient at the bottom boundary, thus even the finest spatial resolution in our simulation was inadequate to capture the too high gradient during the time duration close to the eruption onset (i.e., $t\in [97, 122]$ minutes). The insufficient grid resolution led to the considerable numerical error there, which thus makes the magnetic energy increase higher than the magnetic energy injection in `V3D' simulation. The proper settings of numerical diffusion and grid resolution, along with the improved methods for deriving the photospheric flow will need to be considered in future works for a more accurate data-driven simulation of solar eruptions.

To summarize, our simulation shows that, besides inputting the time series of vector magnetogram, the numerical model of solar corona can be driven to erupt by inputting the time series velocity field at the bottom boundary, in which the driver, i.e., the bottom velocity field is derived from time series of vector magnetograms. The numerical model we established here shows the possibility of driving the evolution of solar corona using different physical variables, which offers a straightforward way for revealing the eruption mechanisms in real events. Thus, such model has a great potentiality in forecasting the onset time as well  as the strength of solar eruptions and  evaluating the quantitative impact on space weather.
~\\

This work is jointly supported by the National Natural Science Foundation of China (NSFC 41731067, 42030204, 42174200), the Fundamental Research Funds for the Central Universities (grant No. HIT.OCEF.2021033), and Shenzhen Science and Technology Program (grant Nos. RCJC20210609104422048 and JCYJ20190806142609035). The computational work was carried out on TianHe-1(A),
National Supercomputer Center in Tianjin, China, and we thank Jun Chen for his informative and helpful discussions.

\end{document}